\documentclass{article}
\usepackage{arxiv}
\usepackage[utf8]{inputenc} 
\usepackage[T1]{fontenc}    
\usepackage{hyperref}       
\usepackage{url}            
\usepackage{booktabs}       
\usepackage{amsfonts}       
\usepackage{nicefrac}       
\usepackage{microtype}      
\usepackage{lipsum}
\usepackage{graphicx}
\graphicspath{ {./images/} }
\usepackage{subfigure}
\usepackage{color, colortbl}

\title{Toward Large-Scale Autonomous Monitoring and Sensing of Underwater Pollutants}

\author{
 Huber Flores \\
  Institute of Computer Science\\
  University of Tartu\\
  Tartu, Estonia \\
  \texttt{huber.flores@ut.ee} \\
   \And
 Naser Hossein Motlagh \\
  Department of Computer Science\\
  University of Helsinki\\
  Helsinki, Finland \\
  \texttt{naser.motlagh@helsinki.fi} \\
  \And
 Agustin Zuniga \\
  Department of Computer Science\\
  University of Helsinki\\
  Helsinki, Finland  \\
  \texttt{agustin.zuniga@helsinki.fi} \\
    \And
  Mohan Liyanage \\
  School of Coumputing and Information\\
  University of Tartu\\
  Tartu, Estonia \\
  \texttt{mohan.liyanage@ut.ee} \\ 
    \And
  Monica Passananti \\
  Department of Chemistry\\
  University of Turin\\
  Turin, Italy \\
  \texttt{monica.passananti@unito.it} \\
    \And
   Sasu Tarkoma \\
  Department of Computer Science\\
  University of Helsinki\\
  Helsinki, Finland  \\
  \texttt{sasu.tarkoma@cs.helsinki.fi} \\
    \And
  Moustafa Youssef \\
  Department of Computer and\\ Systems Engineering\\
  Alexandria University\\
  Alexandria, Egypt \\
  \texttt{moustafa@alexu.edu.eg} \\
    \And
  Petteri Nurmi \\
  Department of Computer Science\\
  University of Helsinki\\
  Helsinki, Finland \\
  \texttt{petteri.nurmi@cs.helsinki.fi} \\
}

\begin{document}
\maketitle
\begin{abstract}
\textit{Marine pollution} is a growing worldwide concern, affecting health of marine ecosystems, human health, climate change, and weather patterns. To reduce underwater pollution, it is critical to have access to accurate information about the extent of marine pollutants as otherwise appropriate countermeasures and cleaning measures cannot be chosen. Currently such information is difficult to acquire as existing monitoring solutions are highly laborious or costly, limited to specific pollutants, and have limited spatial and temporal resolution. In this article, we present a \textit{research vision} of \textit{large-scale autonomous marine pollution monitoring} that uses coordinated groups of autonomous underwater vehicles (AUV)s to monitor extent and characteristics of marine pollutants. We highlight key requirements and reference technologies to establish a research roadmap for realizing this vision. We also address the feasibility of our vision, carrying out controlled experiments that address classification of pollutants and collaborative underwater processing, two key research challenges for our vision.
\end{abstract}


\section{Introduction}

\textit{Marine pollution} is a growing concern worldwide that affects the health of marine ecosystems, weather patterns, climate, and even human health~\cite{thevenon2015plastic}. The main sources of pollution are chemical contaminants and diverse trash, which both result from human activity. Chemical contaminants result from runoff of chemicals into waterways, with particularly agriculture and sewage being major contributors. Trash, in turn, encompasses manufactured products that end up in marine ecosystems due to littering, winds, lacking waste management or human activity. In terms of trash, plastic debris is particularly problematic as plastics are durable and not subject to biological decomposition. This results in steadily growing accumulation of plastics, with estimates suggesting that already in $2014$ over $5$ trillion pieces of plastic were drifting in the oceans~\cite{eriksen2014plastic}. 

Counteracting problems resulting from marine pollution requires efforts to both prevent pollution entering marine areas and to clean up existing pollutants. The former is actively pursued by legislative frameworks, which seek to reduce use and to improve handling of polluting materials, with particularly bans or restrictions on single-use plastics being actively pursued~\cite{schnurr18reducing}. The latter, on the other hand, requires extensive efforts at mapping the extent of pollutants together with costly and laborious cleaning efforts. Currently, obtaining accurate information about the extent of marine pollutants is difficult as existing measurement solutions are highly laborious or costly, limited to specific pollutants, and have limited spatial and temporal resolution. 

\begin{figure*}
  \centering
  \includegraphics[width=1.0\linewidth]{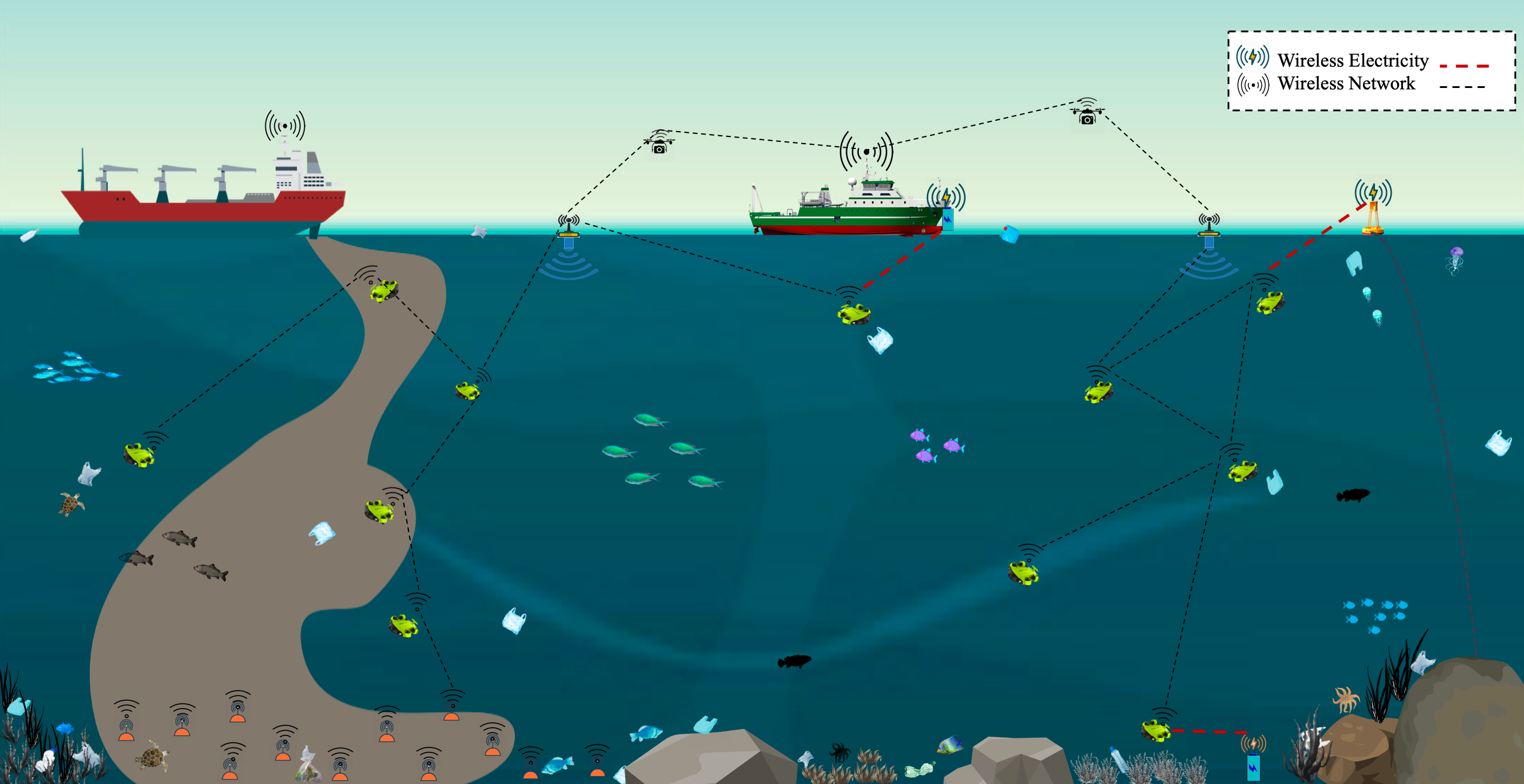}
  \caption{Vision of AUV deployment for plastic detection underwater.}
  \label{fig:underwaterDrone}
\end{figure*}

In this paper, we envision \textit{autonomous marine pollution monitoring} as an approach that can improve current state of pollution monitoring and link with existing research activities on autonomous underwater technologies. Our research vision, illustrated in Figure~\ref{fig:underwaterDrone}, uses \textit{ autonomous underwater vehicles} (AUV)s to identify and classify pollutants in oceans -- or other aquatic environments such as lakes and rivers. In our vision, the AUVs can be responsible for monitoring on their own, or they can link with existing infrastructure. Examples of infrastructure include surface stations, such as instrumented buoys moored at anchor locations or vessels that coordinate and collect results from the AUVs, and fixedly deployed underwater infrastructure, such as sensor networks~\cite{heidemann2012underwater} or energy harvesting stations that use marine or seabed activity to generate power~\cite{rezaei2012energy}. Realizing our vision requires new technological solutions in several research fields. For example, targeting appropriate cleaning activity requires sensing techniques that can identify and classify different pollutants. Such techniques need to take into account characteristics of marine environments as the physical and chemical properties of pollutants change when they are exposed to salt, aquatic organisms, UV radiation and other environmental factors~\cite{passananti2014photoenhanced}. At the same time, sufficient processing power is required to run the sensing models, necessitating new types of collaborative underwater processing infrastructure, highly efficient underwater communication and potentially also new types of energy solutions. To ensure the collected information is of high quality and has sufficient coverage, the AUVs need to coordinate sampling, necessitating advances in underwater localization and AUV coordination. We highlight key research challenges, reflect on current state-of-the-art, and establish a research roadmap for enabling our vision.

We demonstrate the feasibility of our vision by conducting two controlled benchmark experiments that address key research directions of our vision. Our first experiment focused on underwater \textit{pollutant classification} using optical (green light based) sensing. Optical sensing is a promising candidate for pollutant classification as it has small energy footprint and as visible light operates robustly in underwater environments. Existing sensing solutions, such as Fourier transform infrared (FTIR) spectroscopy~\cite{huth2011infrared} or computer imaging~\cite{wang2015aquatic}, are poorly suited for underwater sensing due to being sensitive to environmental characteristics. 
Unlike visible light, infrared has poor propagation characteristics, whereas computer imaging requires heavy processing. Our second experiment focuses on cooperative underwater processing with the aim of supporting \textit{debris identification}. Due to battery limitations, AUVs currently have very limited computational resources making them unsuited for heavy processing. To improve processing, collaborative processing in the form of underwater micro-clouds can be used to increase complexity of processing without burdening individual AUVs. We analyze the influence of distributed processing underwater, and discuss the implications of our results in underwater contexts.  

\section{Requirements}
\label{sec:requirements}

Realising large-scale autonomous marine pollution monitoring requires advances in devices, algorithms, system design, and infrastructure to address limitations of current technology. Next, we discuss key requirements for our vision.

\textbf{Pollution Detection and Classification:} Underwater pollutants come in different forms and shapes, and the optimal countermeasure depends on the pollutant. Indeed, while plastic debris are often highlighted as the main source of pollution, other forms of pollution, such as chemical contaminants (e.g., oil or fertilizer run-off) or microparticles resulting from material degradation can be equally hazardous. Interactions between different pollutants also affect pollutant formation, e.g., chemical contaminants can accelerate oxidation which in turn accelerates plastic degradation resulting in microplastics. To ensuring appropriate cleaning operations can be chosen, a wide range of pollutants thus needs to be captured.

\textbf{Coordinated Sampling:} Covering large underwater areas requires carefully designed sampling strategies and coordination among AUVs. For example, visual surveys rely on transect sampling where information is collected along delineated strips of underwater space. AUVs need to be able to execute transect sampling or related sampling strategies on a large-scale to ensure collected information is maximally useful, and can be integrated with existing modeling techniques. As an example, Lamb et al.~\cite{lamb2018plastic} studied impacts of plastic pollution on reefs through transect samples takes from $159$ different reefs over a three year period. At each reef, the surveyed area was between $60-120$m$^2$ and was sampled using three transects which were separated by $20$ meters (i.e., $1-3$ grids per delineation). Multiple AUVs working in coordination can achieve this goal, offering much larger coverage and enabling more advanced sampling strategies, such as three-dimensional transects. Pollutants are known to reside at different depths depending on characteristics of the pollutant~\cite{thevenon2015plastic}, hence $3$D transects and other advanced strategies are essential for capturing the full range of pollutants. 

\textbf{Coordinated Movement and Processing:} Besides coordinating sampling areas, AUVs should coordinate their movements to ensure they remain in each other's communication range. Coordinated movement also offers additional benefits, such as the potential for using collaborative sensing or processing strategies for improving the recognition of pollutants.

\textbf{Interfaces for Remote Operators:} We envision remote operators to be responsible for specifying bounds of areas that need to be monitored and informing AUVs on potential constraints, e.g., maintaining a minimum distance from sensitive reef vegetation. Within each area, AUVs are then responsible for coordinating sampling and monitoring autonomously. Supporting the interactions between remote operators and AUVs requires designing interfaces for remote operators and systems that can relay the instructions to AUVs operating in the wild. 

\textbf{Fault Tolerance:} Ensuring the monitoring captures useful data requires that AUVs can operate for a sufficiently long time (months or even years without human intervention).  Beyond energy constraints, this requires high reliability from the AUVs as well as the technologies integrated into them. Malfunctions in the operations of the AUVs can result in the vehicles sinking -- potentially making them unrecoverable and part of the pollution problem (AUVs contain metals and plastics). Improving reliability also requires better casing solutions -- both for the AUVs and the sensors equipped into them. With the exception of highly expensive professional-grade submersibles, smaller-scale AUVs can rarely operate beyond sunlight zone (i.e., below $200$ meters). 

\textbf{Pollution Cleaning:} Large-scale pollution monitoring requires harnessing thousands or even millions of AUV to ensure sufficient coverage for monitoring. Deploying a large number of AUVs is only feasible, if they are sufficiently affordable (less than \$10000 ). Current AUVs in this price range are small and have limited capabilities. In terms of cleaning, this means that individual AUVs unlikely can contribute much to cleaning efforts. However, we envision them to take an active role in coordinating surface-based cleaning activities. For example, micro-plastics removal can be operated using specially designer trawlers, whereas larger debris can be collected using surface-based interceptors which can operate using wind and solar power.

\section{Challenges and Enablers}

Enabling fully autonomous marine pollution monitoring is currently difficult due to technological limitations. We next reflect on the current state of technology, and highlight key research challenges in enabling our vision. A summary of the challenges and existing solutions is shown in Figure~\ref{fig:challenges}.


\begin{table}[]
\centering
\caption{Current state of AUV technologies and key challenges and research topics for enabling autonomous underwater pollution monitoring.}
\label{fig:challenges}
\resizebox{.92\textwidth}{!}{%
\begin{tabular}{|l|l|l|l|}
\toprule
\multicolumn{1}{|c|}{} & \multicolumn{1}{c|}{\textbf{State-of-the-Art}} & \multicolumn{1}{c|}{\textbf{\begin{tabular}[c]{@{}c@{}}Key Research \\ Challenges\end{tabular}}} & \multicolumn{1}{c|}{\textbf{\begin{tabular}[c]{@{}c@{}}Emerging \\ Challenges\end{tabular}}} \\ \midrule
\textbf{Sensing} & \begin{tabular}[c]{@{}l@{}}Algorithms and \\ techniques \\ for monitoring\\ individual \\ pollution sources.\end{tabular} & \begin{tabular}[c]{@{}l@{}}Algorithms and \\ techniques for \\ classifying pollution \\ sources and detect \\ multiple pollutants \\ simultaneously. \\ Improved robustness \\ for variations in \\ underwater conditions\end{tabular} & \begin{tabular}[c]{@{}l@{}}Approaches to \\ model internal \\ decomposition and \\ transformation of \\ materials exposed to \\ environmental \\ conditions.\end{tabular} \\ \hline
\textbf{\begin{tabular}[c]{@{}l@{}}Situational \\ Awareness\end{tabular}} & \begin{tabular}[c]{@{}l@{}}Accurate sonar \\ techniques and \\ underwater LIDAR \\ technology. Designs \\ for taller currently \\ bulky, expensive \\ and resource \\ consuming.\end{tabular} & \begin{tabular}[c]{@{}l@{}}Energy-efficient \\ high-resolution \\ situational awareness \\ techniques, including \\ ranging and camera-\\ based solutions.\end{tabular} & \begin{tabular}[c]{@{}l@{}}Coordinated data \\ collection that \\ supports scientific \\ sampling, e.g., belt \\ or grid transects or \\ 3D grid cluster \\ sampling strategies.\end{tabular} \\ \midrule
\textbf{Localization} & \begin{tabular}[c]{@{}l@{}}Ranging based \\ localization and \\ underwater dead \\ reckoning \\ techniques\end{tabular} & \begin{tabular}[c]{@{}l@{}}Positioning schemes \\ for 3D absoulte \\ underwater \\ localization, improved \\ robustness for relative \\ positioning\end{tabular} & \begin{tabular}[c]{@{}l@{}}Hybrid localization \\ solutions that offer \\ relative and absolute \\ positioning, e.g., \\ by interacting with \\ infrastructure \\ residing on the \\ surface or on \\ the sea bottom.\end{tabular} \\ \midrule
\textbf{Communications} & \begin{tabular}[c]{@{}l@{}}Acoustic, optical \\ and electromagnetic \\ underwater \\ communication \\ technologies\end{tabular} & \begin{tabular}[c]{@{}l@{}}High band short-range \\ communication \\ technologies and \\ robust long-range \\ technologies that can \\ interact with other \\ infrastructure.\end{tabular} & \begin{tabular}[c]{@{}l@{}}Improving robustness \\ of communication \\ technologies against \\ water characteristics, \\ such as currents or \\ other water flows, \\ temperature, and \\ salinity.\end{tabular} \\ \hline
\textbf{Design} & \begin{tabular}[c]{@{}l@{}}Fixed AUV \\ designs that are \\ tightly sealed \\ and difficult to \\ expand\end{tabular} & \begin{tabular}[c]{@{}l@{}}Designs that allow \\ additional modules, \\ such as sensing or \\ external processing \\ units, to be integrated \\ into AUV's.\end{tabular} & \begin{tabular}[c]{@{}l@{}}Lightweight casing \\ materials that offer \\ water and pressure \\ protection, but do not \\ hamper sensing or \\ communication \\ functionality.\end{tabular} \\ \midrule
\textbf{\begin{tabular}[c]{@{}l@{}}Power and \\ Operational \\ Time\end{tabular}} & \begin{tabular}[c]{@{}l@{}}Battery-based \\ AUV's with short \\ operating time. \\ Low-efficiency \\ energy harvesting \\ solutions.\end{tabular} & \begin{tabular}[c]{@{}l@{}}Novel power solutions, \\ including wireless \\ charging stations, \\ underwater energy \\ harvesting and other \\ solutions.\end{tabular} & \begin{tabular}[c]{@{}l@{}}Collaborative \\ underwater processing \\ and offloading for \\ resource augmentation \\ and improved energy \\ efficiency.\end{tabular} \\ \bottomrule
\end{tabular}%
}
\end{table}

\textbf{Sensing:} Accurately detecting and classifying different pollutants requires new types of sensing solutions and systems that can effectively combine sensing modalities. Currently, chemical contaminants can be identified using underwater mass spectrometers~\cite{farrell2005chemical}. These systems, however, are not suitable for large-scale monitoring as underwater mass spectrometers have limited operating time, are difficult to integrate to submersibles due to need for a vacuum environment for taking measurements, and suffer from high power draw. Overcoming these challenges requires new power solutions and sensing approaches that can detect the presence of chemical contaminants, limiting the use of spectrometers to contaminated areas. For debris, a common approach for material detection is to rely on Fourier transform infrared (FTIR) spectroscopy~\cite{huth2011infrared}. Water heavily absorbs infrared light, making this approach ill-suited for underwater use. FTIR also requires special microscopes which are costly, and difficult to integrate as part of AUVs. Increasing the scale of underwater monitoring thus requires novel low-cost sensing techniques that have low power draw and can operate effectively underwater. Another challenge for debris monitoring is that the materials undergo chemical and physical changes as a result of exposure to UV radiation and aquatic organisms~\cite{passananti2014photoenhanced}. In practice, it is likely that a combination of sensing techniques needs to be adopted. For example, many AUVs have cameras which can be used to detect the presence of debris~\cite{wang2015aquatic} and other sensing techniques can then be used to provide a more fine-grained classification (e.g., which type of plastic). Enabling large-scale autonomous pollution monitoring thus requires novel sensing solutions that can operate with low energy footprint, are able to classify different types of pollutants, and work robustly against changes in the physical or chemical composition of materials. 

\textbf{Situational Awareness:} Ensuring AUVs coordinate their operations is critical for large-scale autonomous underwater pollution tracking. Autonomy and coordination have been extensively studied in aerial and ground operations~\cite{motlagh2016low}, but existing techniques cannot be directly adapted in underwater environments, requiring development of new solutions and research on adapting existing techniques into underwater environments. AUVs need to have sufficient degree of situational awareness to be able to operate effectively. For example, AUVs need information about obstacles  (fish or other aquatic organisms, rocks, and other types of obstacles). In most aquatic environments, simple sonar-based solutions are likely sufficient. However, sensitive environments, such as reefs, likely require more precise information. Current high-precision solutions, such as underwater LIDARs, are too expensive, bulky and power consuming for widespread usage, necessitating research on new techniques and miniaturization of existing technologies. There is also a need for effective orchestration mechanisms that allow coordinating AUV operations and interface with remote operators. 

\textbf{Underwater Localization:} Coordination requires AUVs to be aware of their position relative to other devices. For targeting sampling at the correct areas, it is also critical that AUVs are aware of their global (three dimensional) location. Currently no equivalent of GPS exists for underwater environments with inertial and relative positioning being the best options. Thus, there is a need for improved underwater localization and/or hybrid solutions that rely on relative positioning and surface-based stations.

\textbf{Communication:} Coordination between AUVs and other infrastructure is reliant on robust and sufficiently high bandwidth communication links. Underwater communications are among the most widely studied aspects pertaining to our vision with acoustic, optical, ultrasonic, electromagnetic and radio frequency based techniques being examples of solutions that have been proposed~\cite{che2010re}. Current technologies, however, are insufficiently mature for unsupervised operation as they suffer from limited range or bandwidth or are sensitive to environmental characteristics such as salinity, water temperature or currents. We envision AUVs to actively share information, e.g., sharing environmental information or offloading images for improved debris recognition. This requires communication links with sufficiently high bandwidth. In practice, realising our vision likely requires a combination of long and short-range communication techniques. For example, acoustic communications can be used to connect with surface infrastructure (e.g., instrumented buoys or ships) whereas optical communication offers sufficient range and bandwidth for local connectivity. Optical communication has the added benefit of being able to support simultaneous sensing and communication. Both optical and acoustic communications are sensitive to characteristics of underwater environments and thus there is a need for improving their robustness or for developing alternative communication mediums. 

\textbf{Design:} Sensing and computing units traditionally are averse to water, requiring  watertight casing when integrated into AUVs. This makes it difficult to extend or adapt operations of AUVs. Indeed, most AUVs being designed for a single dedicated task, such as estimating water velocity~\cite{eriksen2001seaglider} or debris monitoring~\cite{wang2015aquatic}. Realising our vision requires improved AUV designs that are modular and extensible, allowing different types of sensors, additional computing units, and even power sources to be attached to AUVs. At the same times, AUVs need to offer better programming interfaces and SDKs that allow adapting them to different operations. Finally, ensuring sustainability  requires research on new types of materials that are  durable and harmless to the environment in case accidents happen. As sensing and communication units need encasing, the materials should also be such that they do not degrade sensing or communication performance. 

\textbf{Power and Operational Time:} Commercial off-the-shelf AUVs have limited operational time, surviving at most $24$ hours between recharges. For example, PowerRay, one of the most popular commercial underwater drones, reportedly has a $4$ hour operational time while submerged. To avoid loss, the internal processing units of AUVs commonly integrate routines that estimate travel times and prevent travel if there is not enough power for a return trip. Large-scale monitoring thus requires designated station points where AUVs can be moored and charged. Optimally charging would happen without human intervention, e.g., using buoys that harvest solar energy or seabed sensors harvesting energy from currents and seabed motions~\cite{rezaei2012energy}. This is difficult to achieve as station points require additional deployment and maintenance. Besides novel power solutions, operational time can be improved by designing mechanisms for energy-efficient operations. For example, image processing operations can be performed collaboratively or offloaded to surface-based infrastructure. Indeed, in the near future it may be possible for AUVs to link with data centers that are located on the surface~\cite{cutler2017dunking}.

\section{Feasibility Experiments}

We address feasibility of our vision through two benchmark experiments addressing two of the key research challenges described in the previous section. Our first experiment targets energy-efficient classification of debris based pollutants using optical green light sensing, whereas our second experiment examines the feasibility of using commercial-off-the-shelf devices for augmenting the processing resources of AUVs. We next briefly summarize the experimental setup of our two benchmarks. In the experiments we rely on measurements from controlled testbeds to have better control over experimental variables. We have also separately verified the feasibility of these technologies in AUV operations by integrating the sensors onto a PowerRay underwater drone. 

\begin{figure}
    \begin{center}
        \includegraphics[width=10.5cm]{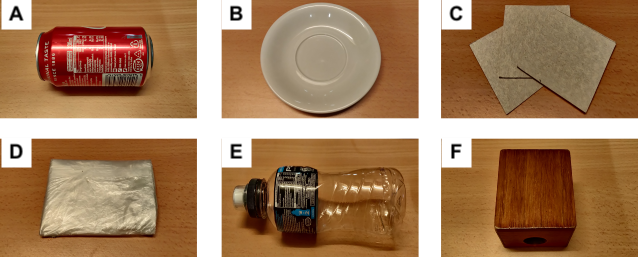}
        \caption{Debris materials. A. Aluminium can, B. Ceramic plate, C. Paperboard, D. Plastic bag, E. Plastic bottle, F. Wooden toy.}
        \label{fig:Materials}
    \end{center}
\end{figure}

\subsection{Underwater Material Sensing}
\label{ssec:materialsetup}

Having a complete view of pollutant types and characteristics is necessary for targeting appropriate cleaning actions and estimating the severity of the current pollution situation. Current monitoring solutions are unable to achieve this as they can only detect the presence of pollutants without being able to classify them. Our first experiment demonstrates that optical sensors available on commercial-off-the-shelf devices can be used to perform coarse-grained debris classification. We focus on optical sensing as the sensors are inexpensive, have low power draw, and are capable of operating normally underwater.

\textbf{Apparatus:} We perform our experiment using a commercial off-the-shelf smartwatch (Samsung Gear S3 Frontier\footnote{https://www.samsung.com/global/galaxy/gear-s3/}) which integrates two green light LED lights and a photo-receptor. We focus on green light due to its short wavelength, which makes it excellent at penetrating water. 

\textbf{Materials:} We test against common everyday objects with different materials. The objects are shown in Figure~\ref{fig:Materials} and they were chosen as representative examples of common types of underwater debris. We consider the following objects: a snack box (paperboard, PAP21), a plastic bag (high-density polyethylene HDPE), a plastic bottle (polyethylene terephthalate PET),  a soft drink can (aluminium, ALU 41), a small ceramic plate (feldspar) and a wooden toy box (solid walnut oak). 

\begin{figure}
    \begin{center}
        \subfigure[Underwater material sensing setup.]{
           \includegraphics[width=6.4cm]{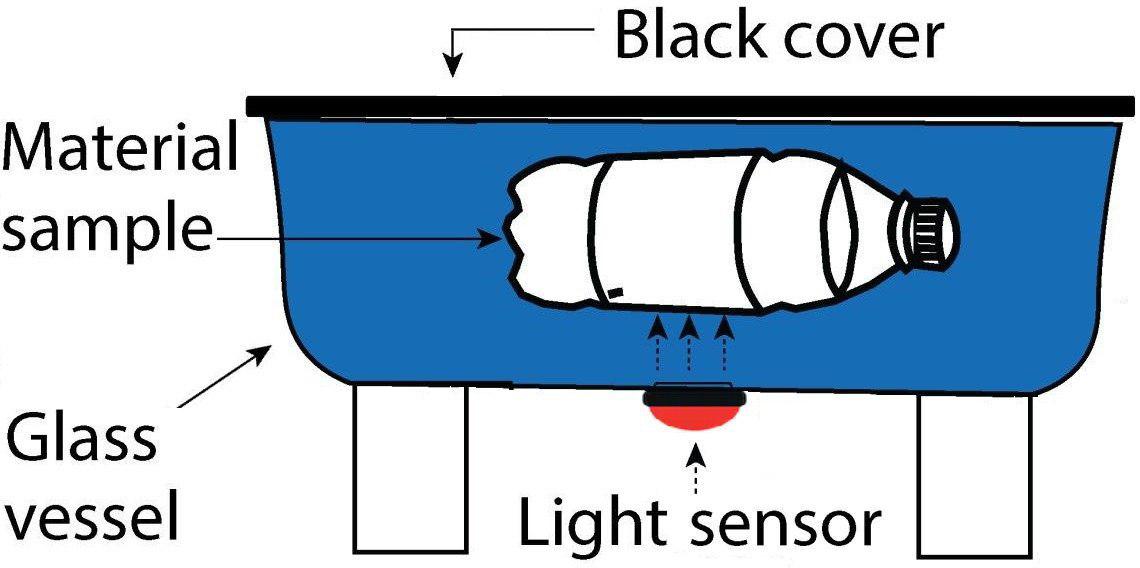}
            \label{fig:materialunderwaterTtested}
        }
        \subfigure[
        Micro-cloud deployment.
        ]{
            \includegraphics[width=5.2cm]{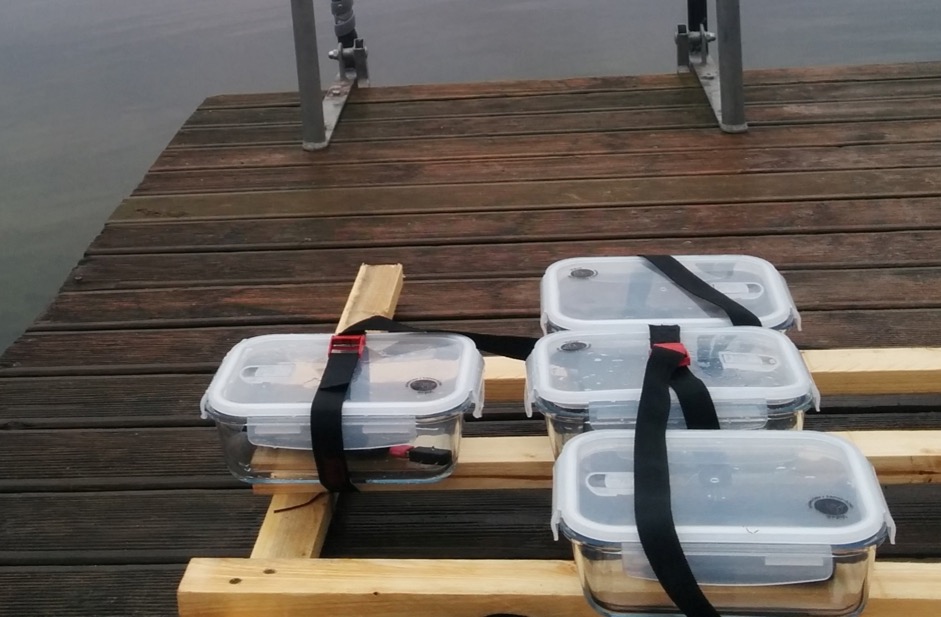}
            \label{fig:microcloud-testbed}
        }
        \caption{Our controlled testbed for underwater material sensing and collaborative processing.}
        \label{fig:ExperimentalSetup}
    \end{center}
\end{figure}

\textbf{Setup:} We place each object in turn into a glass container covered with a non-reflective (black) lid. The smartwatch is taped outside the container, directly below the measured object; see Figure~\ref{fig:materialunderwaterTtested}. For each object, we perform $4$ sets of measurements with each set consisting of $6$ repetitions. Each repetition contains light intensity measurements sampled with $100$Hz frequency over a $90$ second period. The four sets were divided into a $2 \times 2$ design with sensing medium (air cf.~water) and luminosity (ambient cf.~darkness) as the experimental conditions. The two conditions for sensing medium refer to whether the container was empty or filled with water. The former emulates having a water-proof casing on the surface, whereas the latter emulates detection in an underwater environment. The luminosity conditions correspond to a case where the container is unobstructed, emulating case where ambient light seeps into water, and darkness, emulating direct contact between the sensor and the object. The average strength of ambient light was measured as $15.5$lx. Darkness was achieved by covering the container with a cardboard box. We separately verified that luminosity inside the box was zero.

\subsection{Underwater Distributed Processing}

Scaling up autonomous pollution monitoring requires participating AUVs to be affordable yet sufficiently powerful to carry out the required operations. Currently this is not the case, with affordable AUVs having limited computational power and short operational time. For example, most AUVs integrate a Raspberry PI ($1.3$Ghz CPU) or equivalent micro-controller, and have an operational time of $1.5-4$ hours without recharging. In our second experiment, we demonstrate how micro-clouds formed from smart devices can be used to augment the computational capabilities of the AUVs and to prolong their operational time. 

\textbf{Apparatus:} We perform our experiment using a micro-cloud consisting of four LG Nexus $5$ smartphones. Each smartphone is placed inside a sealed glass container, which in turn is attached on a wooden structure equidistant from each other; see Figure~\ref{fig:microcloud-testbed}. During the experiment we submerge the casings to evaluate underwater functionality.

\textbf{Task:} As the experimental task we consider object recognition from a video-feed. We chose this task as it emulates the computational needs of vision-based underwater debris recognition. As the video feed, we consider a set of $50$ images ($224x224$ resolution) taken from ImageNet. For recognition, we use a pre-trained and quantized mobilenet model $(\_v1\_1.0\_224)$ which is deployed on each smartphone.

\textbf{Prototype:} We implemented a proof-of-concept prototype Android app that was deployed on the test devices. The app uses WiFi to connect to the other participating devices. One device is randomly chosen as a master that initiates computing on the other devices, which act as workers. To emulate a real-time video feed, the master processes the images sequentially, sending them one-by-one to a worker node that is chosen in a round robin fashion. Once a worker finishes its task, it sends the results back to the master. 

\textbf{Metrics:} As the evaluation metrics we consider task completion rate, i.e., the number of frames processed by slaves, and task success rate, i.e., the number of returned frames that are successfully received by master. As our focus is on assessing feasibility of underwater offloading and as we use a pre-trained model, evaluating object recognition performance was not meaningful and is omitted. The benefits of running object recognition on smart device micro-clouds have been demonstrated in our previous work~\cite{lagerspetz2019pervasive} and hence we focus on task completion and success rates.

\section{Results}

\subsection{Underwater Material Sensing}

We first use statistical significance testing to verify that the green light intensity measurements of different materials indeed have sufficient variation to support debris identification. Kruskal-Wallis test showed significant differences for all conditions:  air - ambient ($\chi^2 = 6900$, $\eta^2 = 0.95$, p$< .001$), water - ambient ($\chi^2 = 7003$, $\eta^2 = 0.97$, p$<.001$), air - darkness ($\chi^2 = 7005$, $\eta^2 = 0.97$, p$< .001$) and water - darkness ($\chi^2 = 7004$, $\eta^2 = 0.97$, p$< .001$). Here air and water refer to the conditions for sensing medium, i.e., having the glass container empty or filled with water to emulate underwater environment, whereas ambient and darkness refer to the two luminosity conditions; see Section~\ref{ssec:materialsetup}. Posthoc comparisons (Dunn-Bonferroni) verified that the differences for objects also were statistically significant in all experiment conditions. 

We next demonstrate that optical sensing can provide a coarse-grained classification of debris by using the light intensity measurements to run classification experiments. We test using two simple classifiers, a random forest model and a $k$-nearest neighbor classifier, and under different evaluation scenarios. We focus on simple classifiers as the models that are deployed on AUVs need to be simple to run to ensure they are as energy-efficient as possible. The results of our experiments are shown in Table~\ref{tab:classification}. We have separately tested how change in sensing medium (air cf. water) and in luminosity conditions affects performance. When all four test conditions are included, the classification accuracy is around $70\%$. The best classification performance, slightly over $80\%$, is obtained when both testing and training measurements are from environments with similar luminosity values (rows ambient and darkness in the table). Changes in sensing medium have some effect on performance, but these are not as pronounced as in the case of luminosity changes. Overall, the results demonstrate that optical sensing can be used to provide a coarse-grained classification of different pollutants, as long as the training samples provided by the machine learning algorithms are sufficiently similar to the environment where the AUVs operate.

\begin{table}
\centering
\caption{Classification accuracy in different experimental conditions.}
\label{tab:classification}
\resizebox{0.6\textwidth}{!}{%
\begin{tabular}{llll}
\toprule
\textbf{Cross Validation test} & \textbf{k-NN} & \textbf{Random forest} & \textbf{Average} \\
\midrule
All conditions 6-folds & \multicolumn{1}{c}{70.8} & \multicolumn{1}{c}{72.9} & \multicolumn{1}{c}{\textbf{71.9}}  \\
Ambient 6-folds & \multicolumn{1}{c}{81.2} & \multicolumn{1}{c}{82.2} & \multicolumn{1}{c}{\textbf{81.7}}  \\
Darkness 6-folds & \multicolumn{1}{c}{81.2} & \multicolumn{1}{c}{77.1} & \multicolumn{1}{c}{\textbf{79.2}}  \\
Air 6-folds & \multicolumn{1}{c}{76.0} & \multicolumn{1}{c}{71.9}  & \multicolumn{1}{c}{\textbf{74.0}} \\
Underwater 6-folds & \multicolumn{1}{c}{66.7} & \multicolumn{1}{c}{64.6} & \multicolumn{1}{c}{\textbf{72.5}}  \\
\midrule
\textbf{Average} & \multicolumn{1}{c}{\textbf{74.8}} & \multicolumn{1}{c}{\textbf{73.5}} & \multicolumn{1}{c}{\textbf{74.1}}  \\
\bottomrule 
\end{tabular}%
}
\end{table}

\subsection{Underwater Distributed Processing}

\begin{figure}
\centering
\includegraphics[width=0.5\textwidth]{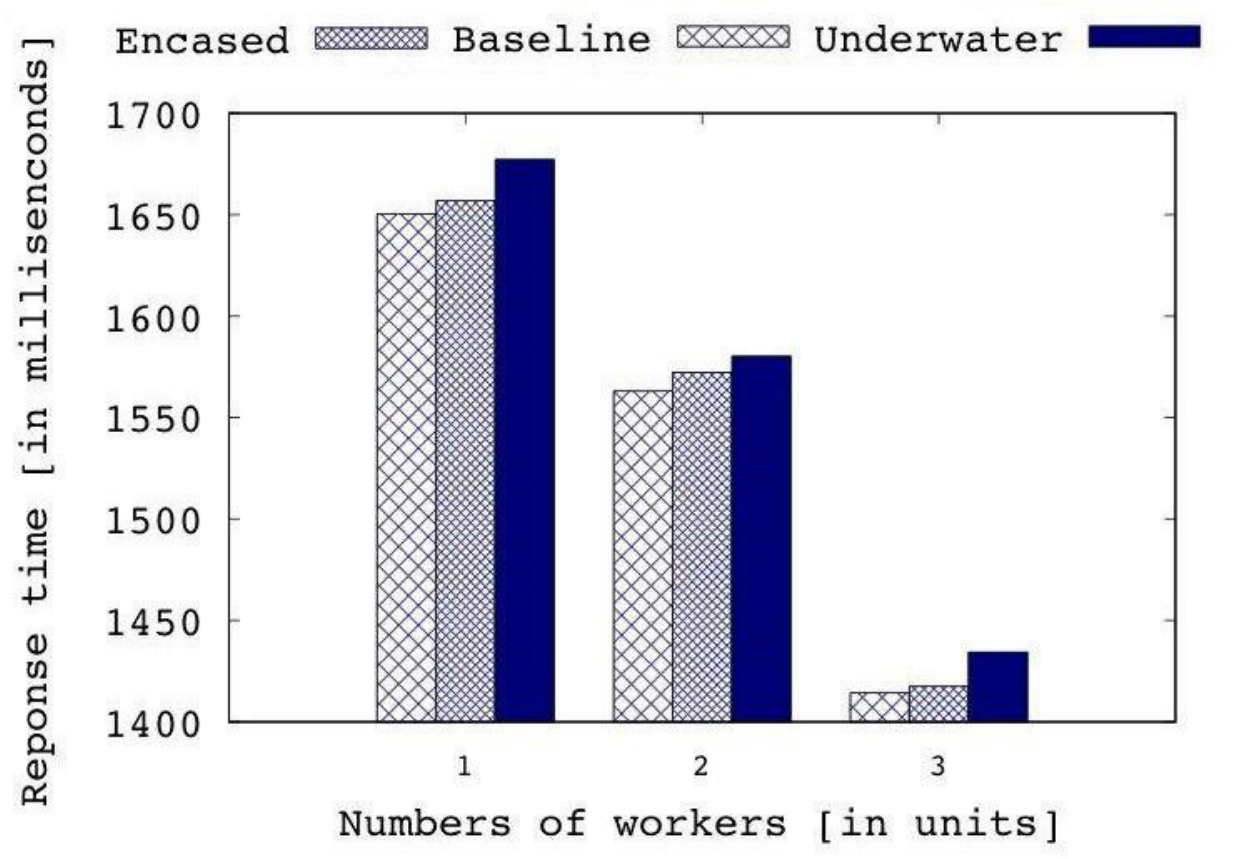} 
\caption{Collaborative processing with waterproof micro-clouds.}
\label{fig:processingUnderwater}
\end{figure}

We first consider the processing time of the devices participating in the micro-cloud. Results of this evaluation are shown in Figure~\ref{fig:processingUnderwater}. As expected, including more workers decreases processing time. Even with one worker device, it is possible to reach $30$fps processing frequency, demonstrating that underwater micro-clouds could be used to support image-based debris recognition. Adding more workers would allow using faster rate of video or higher image resolution. We separately assessed how the encasing or submerging of the devices affects processing. From the figure we can observe that both have some effect on the processing time, but that this effect is overall negligible (approximately $5$ms for the encasing and approximately $20$ms for submerging). 

We next consider the task completion rate by measuring the fraction of object recognition results that are successfully returned to the master. When the devices operate above the surface, the success rate is $100\%$ also when the devices are encased in the glass container. Once the devices are submerged, the success rate decreases -- as could be expected. When devices are close to each other (within $7$cm), success rate remains at $100\%$. However, once distance exceeds this, the performance drops, first to $70\%$ at around $10cm$ distance, and then to $0\%$ beyond $10$ cm distance. Our results thus suggest that underwater collaborative processing itself is feasible with reasonably inexpensive components, but there is a need for short-range communications solutions that can operate efficiently underwater. Alternatively, devices could operate inside a single container, in which case wireless communications would be sufficient. However, in this case the weight of the container easily becomes an issue for operating the AUV.

\section{Discussion}

Our discussion thus far has focused on technological challenges in enabling large-scale autonomous marine pollution monitoring. Beyond these, there are numerous other challenges, e.g., related to the operations and the overall service ecosystem. Below we briefly highlight some of these aspects:

\textbf{Marine Natural Hazards:} AUV technology operating in aquatic ecosystems can be perceived as invasive by wildlife and be subjected to attacks or other unpredictable behaviors. These operations may be harmful to aquatic wildlife, or result in damage to AUVs. Overcoming this issue requires better understanding of different AUV designs and how they are perceived by different species, as well as algorithms that allow AUVs to adapt their behavior to minimize disruptions to aquatic wildlife. 

\textbf{Further Issues:} We have sought to highlight key research challenges related to technology and to demonstrate the feasibility of underwater pollution monitoring. In practice, several other factors affect the adoption of AUVs. For instance, operational regulations and legislation on operating AUVs, especially in marine areas intersecting borders of multiple countries. Other factors include integration of AUVs with legacy technologies, such as already deployed underwater sensor networks, and fabrication of suitable materials for manufacturing of AUVs, e.g., using 3D printing.

\textbf{Marine Data:} The results of our feasibility experiments demonstrated how combinations of different existing technologies could be adopted for improving resolution of underwater pollution monitoring. In particular, collaborative underwater processing would enable image-based debris recognition, whereas optical sensing could be used to support more fine-grained classification into different materials. Improving the maturity of these technologies, and supporting the development of new sensing modalities, would benefit from large-scale datasets that can be used to train and test different solutions.

\textbf{Value Added Services and Stakeholders:} Sustainability of marine life and water resources is a primary concern for governmental institutions across the globe, and aquaculture industries in open waters. Thus, these are primary entities that will foster large-scale and permanent deployments of AUVs. Since AUV deployment is relatively easy and flexible, other groups can also benefit indirectly from spontaneous deployments to monitor emergency events underwater and to offer other monitoring services, e.g., oil spills or fishery monitoring. Supporting the requirements of external stakeholders requires new types of service orchestration schemes that allow easily adapting AUVs for different monitoring tasks, data visualizations that can provide actionable feedback on detection results, and even privacy and security solutions that help preserve data confidentiality in case of unauthorized access, e.g., due to previous loss of equipment.

\bibliographystyle{unsrt}


\end{document}